\documentclass[twocolumn]{aastex63}

\begin{document}

\newcommand{\kms}{km~s$^{-1}$}	\newcommand{\cms}{cm~s$^{-2}$}
\newcommand{\msun}{$M_{\odot}$} \newcommand{\rsun}{$R_{\odot}$} 
\newcommand{\teff}{$T_{\rm eff}$} \newcommand{\logg}{$\log{g}$} 
\newcommand{\mas}{mas~yr$^{-1}$}


\title{ A 1201~s Orbital Period Detached Binary:  the First Double 
Helium Core White Dwarf LISA Verification Binary}

\author[0000-0002-4462-2341]{Warren R.\ Brown} \affiliation{Smithsonian 
Astrophysical Observatory, 60 Garden Street, Cambridge, MA 02138 USA}

\author[0000-0001-6098-2235]{Mukremin Kilic} \affiliation{Homer L. Dodge Department 
of Physics and Astronomy, University of Oklahoma, 440 W. Brooks St., Norman, OK, 
73019 USA}

\author[0000-0002-2384-1326]{A.\ B{\'e}dard} \affiliation{D{\'e}partement de 
Physique, Universit{\'e} de Montr{\'e}al, C.P. 6128, Succ. Centre-Ville, 
Montr{\'e}al, Quebec H3C 3J7, Canada}

\author[0000-0002-9878-1647]{Alekzander Kosakowski} \affiliation{Homer L. Dodge 
Department of Physics and Astronomy, University of Oklahoma, 440 W. Brooks St., 
Norman, OK, 73019 USA}

\author[0000-0003-2368-345X]{P.\ Bergeron} \affiliation{D{\'e}partement de Physique, 
Universit{\'e} de Montr{\'e}al, C.P. 6128, Succ. Centre-Ville, Montr{\'e}al, Quebec 
H3C 3J7, Canada}

\shortauthors{Brown et al.}
\shorttitle{ J2322+0509: the First He+He White Dwarf LISA Binary}

\begin{abstract}

	We report the discovery of a 1201~s orbital period binary, the third 
shortest-period detached binary known.  SDSS J232230.20+050942.06 contains two 
He-core white dwarfs orbiting with a $27\arcdeg$ inclination.  Located 
0.76 kpc from the Sun, the binary has an estimated LISA 4-yr signal-to-noise 
ratio of 40.  J2322+0509 is the first He+He white dwarf LISA verification 
binary, a source class that is predicted to account for one-third of resolved 
LISA ultra-compact binary detections.

\end{abstract}


\section{INTRODUCTION}

	White dwarf (WD) binaries promise to be among the most scientifically rich 
``multi-messenger'' sources that can be observed with both light and gravitational 
waves (GW).  ESA and NASA are building a space-based GW observatory called the Laser 
Interferometer Space Antenna \citep[LISA,][]{amaro17} that will measure mHz GW 
frequencies.  Astrophysically, this is the realm of $<$1~hr orbital period 
double-degenerate binaries.  WD+WD binaries are expected to be the most prolific 
LISA source \citep{nelemans01a, korol17}.  Importantly, WDs emit light and we can 
observe them now.

	Here, we report the discovery of the 1201~s orbital period binary SDSS 
J232230.20+050942.06 (hereafter J2322+0509).  This is the third shortest-period 
detached binary after J0651+2844 \citep{brown11b} and ZTF J1539+5027 
\citep{burdge19}.  While similar in period to J0935+4411 \citep{kilic14} and PTF 
J0533+0209 \citep{burdge19b}, J2322+0509 has a face-on orientation $i=27\arcdeg$.  
It has no detectable light curve or binary astrometric signal; only time-series 
spectroscopy is able to detect its period.

	We combine spectroscopy, photometry, and astrometry to characterize the 
system.  J2322+0509 is a single-lined spectroscopic binary with a 19,000~K WD moving 
with a velocity semi-amplitude of $k=149$~\kms.  Multi-passband photometry shows 
that a cooler companion contributes an additional 15\% extra light at red 
wavelengths.  Astrometry sets an absolute distance to the system, and places a 
direct constraint on the WD radii.  We perform a joint analysis and conclude that 
J2322+0509 is an approximately equal mass, double-degenerate binary containing a 
0.27~\msun\ DA WD and a 0.24~\msun\ DC WD.

	J2322+0509 is thus the first double He-core (He+He) WD LISA verification 
binary.  Models predict this source class will account for about one-third of 
resolved LISA ultra-compact binary detections \citep{lamberts19}.  LISA will detect 
J2322+0509 with an estimated signal-to-noise ratio SNR~$\simeq40$ in 4 years of 
operation.

\section{Observations}

	We targeted J2322+0509 on the basis of its Gaia parallax and Sloan Digital 
Sky Survey (SDSS) colors.  Following the release of Gaia Data Release 2 
\citep{gaia18}, we searched the Gaia catalog for extremely low mass WD candidates 
that might be missing from the ELM Survey \citep{brown20}.  We selected candidates 
with de-reddened SDSS color $-0.40<(g-r)_0<-0.10$ mag (approximately $20,000>T_{\rm 
eff}>10,000$~K) and with Gaia parallax consistent with a $\sim$0.1 \rsun\ low mass 
WD.  J2322+0509, at the blue edge of the sample, is an interesting result.  We 
present our observations followed by our analysis.  Measured and derived values are 
summarized in Table \ref{tab:param}.

\begin{figure} 
 \includegraphics[width=3.4in, clip, bb=10 180 570 695]{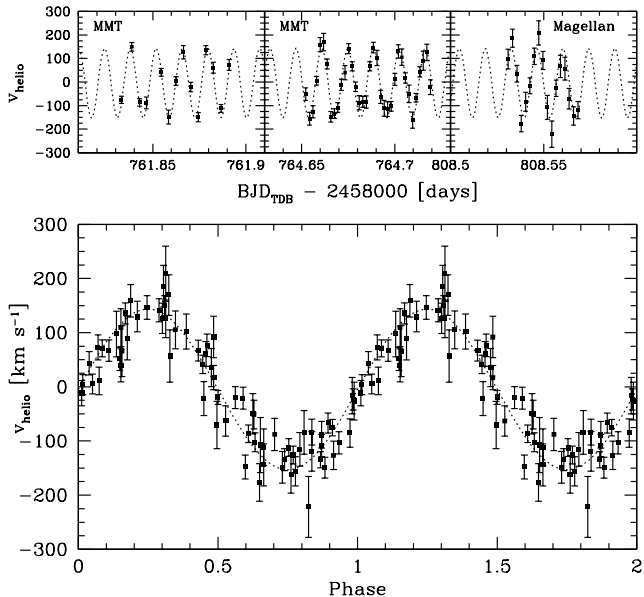}
 \caption{ \label{fig:rv}
	Radial velocities phased to the best-fit orbital solution.  The measurements 
are available in the electronic Journal as the Data behind the Figure.}
 \end{figure}

\subsection{Spectroscopy}

	We obtained an exploratory spectrum of J2322+0509 on UT 2018 December 9 to 
determine its stellar nature.  We used the Blue Channel spectrograph on the 6.5m MMT 
telescope with the 832 l~mm$^{-1}$ grating in 2nd order, giving 1.0 \AA\ resolution.  
J2322+0509 has the hydrogen Balmer line spectrum of a DA WD.  We fit pure hydrogen 
stellar atmosphere models \citep{gianninas11, gianninas14, gianninas15} and 
determined that J2322+0509 is a $\log{g}\simeq7$ WD.  WDs with such gravities are 
commonly found in ultra-compact binaries, because the Universe is not old enough to 
evolve a low mass WD through single-star evolution \citep[e.g.][]{iben90, marsh95}.

	The following year we obtained time-series MMT Blue Channel spectroscopy to 
test for binary orbital motion.  We observed low-amplitude radial velocity 
variability, but were unable to determine a period until we took back-to-back 
exposures on UT 2019 October 5.  On UT 2019 October 8, we observed J2322+0509 with a 
lower 2.0 \AA\ resolution set-up, using the MMT Blue Channel 800 l~mm$^{-1}$ 
grating, so that we could better time-resolve the orbital period (150~s = 1/8 
orbital phase) at the cost of lower velocity precision.  Figure \ref{fig:rv} plots 
the measurements.

	Finally, on UT 2019 November 20, we obtained 1 hr of MagE spectroscopy on 
the 6.5m Magellan Baade telescope to test for spectral lines from a companion in the 
red end of the spectrum.  We used the 0.85 arcsec slit and 180~s exposures.  We 
confirmed the radial velocity parameters of the primary WD, but phase-folding the 
spectra reveals no evidence of spectral lines from the secondary.

\subsection{Photometry}

\begin{figure} 
 \includegraphics[width=3.25in, clip, bb=10 155 570 695]{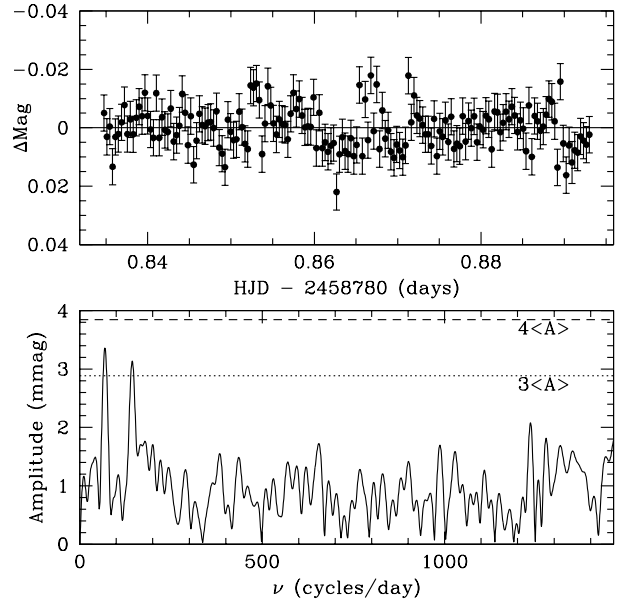}
 \caption{ \label{fig:lc}
	Optical light curve (upper panel) and its Fourier transform (lower panel).  
The absence of variability above 4$<$A$>$ = 0.35\% sets an upper limit on 
inclination $i<50\arcdeg$. } 
\end{figure}

	We obtained time-series photometry of J2322+0509 using the 8-m Gemini North 
telescope with the Gemini Multi-Object Spectrograph (GMOS) on UT 2019 October 24 as 
part of the Director's Discretionary Time program GN-2019B-DD-201. We obtained 170 
$\times$ 10~s exposures through an SDSS-$g$ filter over 84~min. To reduce the 
read-out time, we binned the chip by 4$\times$4, which resulted in a plate scale of 
$0.3\arcsec$ pixel$^{-1}$. Conditions were photometric with 0$\farcs$5 seeing.  
Figure \ref{fig:lc} shows the lightcurve and its Fourier transform.

	We observe no significant variability at the 4$<$A$>$ = 3.8 millimag 
(0.35\%) level.  Interestingly, the highest peak in the Fourier transform -- 
$3.1\pm0.9$ millimag amplitude at $69\pm3$ cycles day$^{-1}$ -- is consistent with 
the radial velocity period, however it is only marginally significant.  The 
predicted amplitude of the relativistic beaming effect is only 0.1\% 
\citep{shporer10}, below our detection threshold.  No eclipses or ellipsoidal 
variation are detected.

\begin{figure*} 
 \centerline{ \includegraphics[height=7.0in, angle=270, clip, bb=140 60 480 740]{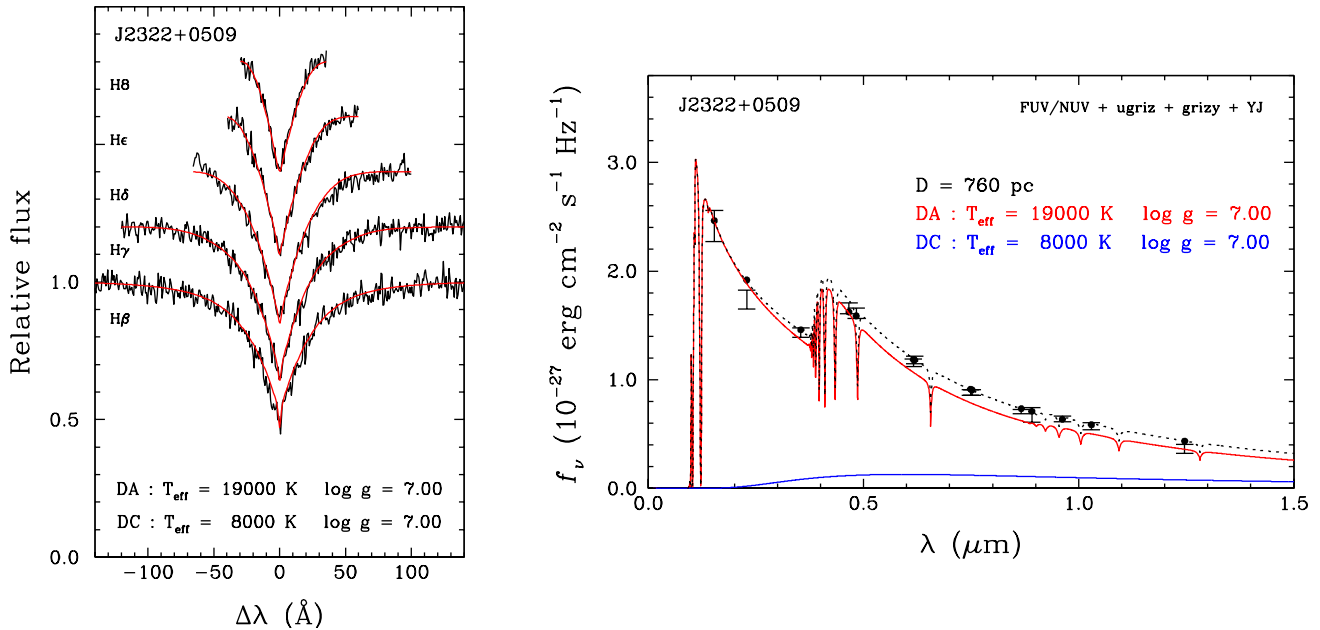} }
 \caption{ \label{fig:joint}
	Joint fit to the spectrum (left panel) and photometry (right panel) for the 
best DA+DC model.  The left panel shows the synthetic model (red) overplotted on the 
observed MMT spectrum (black).  The right panel shows the synthetic fluxes (filled 
circles) and observed fluxes (error bars).  The red and blue lines show the 
contribution of each WD to the total monochromatic model flux, displayed as the 
black dotted line.}
 \end{figure*}

\section{Analysis}

	J2322+0509 is a single-lined spectroscopic binary.  We solve for its radial 
velocity orbital parameters following the same procedure used in previous ELM Survey 
papers \citep{brown20}.  We start by cross-correlating the full spectra with a 
summed, rest-frame template of the target.  Comparing the independent MMT and 
Magellan datasets suggests our zero-point accuracy is 2 \kms.  We then minimize 
$\chi^2$ for a circular orbit solution, accounting for phase smearing given the 
exposure times.  We estimate internal uncertainties by bootstrap re-sampling the 
measurements.  The best-fit radial velocity solution of the combined MMT and 
Magellan datasets is $P=1201.4\pm5.9$~s, $k=148.6\pm6.3$ \kms, and 
$\gamma=-4.8\pm4.2$ \kms\ (see Figure \ref{fig:rv}).

	We fit pure hydrogen (DA WD) stellar atmosphere models to the summed, 
rest-frame MMT and Magellan spectra.  The \teff\ and $\log{g}$ values from the 
independent MMT and Magellan datasets are consistent within their errors.  The 
weighted mean values are $T_{\rm eff,spec} = 19,160 \pm 270$~K, $\log{g}_{\rm spec} 
= 7.17 \pm 0.04$ dex.

	High-order Balmer lines are sensitive to surface gravity \citep{tremblay09} 
and allow us to constrain the DA WD mass given a WD mass-radius relation.  We adopt 
the He-core WD models of \citet{althaus13} appropriate for this object.  
Interpolating the stellar atmosphere values with their errors through the models 
yields $M_{\rm DA, spec} = 0.34 \pm 0.02$ \msun.  The corresponding $g$-band 
absolute magnitude, $+9.27 \pm 0.18$ mag, allows us to compute its heliocentric 
distance if the light of the DA WD dominates the light of the (de-reddened) 
$g_0=18.351 \pm 0.018$ mag binary.

	However, we find evidence for additional light from the companion star in 
the broadband spectral energy distribution.  If we normalize the DA WD model to the 
SDSS $u$-band magnitude, the $g$- and $r$-bands are $15\pm2$\% too bright.  
Accounting for this excess $g$-band light, our spectroscopic parameters imply the DA 
WD is $d_{\rm spec}=0.72\pm0.07$ kpc distant.

	Gaia parallax provides an independent but less precise measure of 
heliocentric distance.  Adopting the Gaia DR2 parallax zero-point of $-0.029$~mas 
\citep{lindegren18}, J2322+0509's $1.287 \pm 0.283$ mas parallax corresponds to 
$d_{\rm plx}=0.76 \pm 0.17$ kpc.  The spectroscopic and parallax distances are in 
perfect agreement.

	The absence of a significant reflection effect constrains the orientation of 
the binary.  We use JKTEBOP \citep{southworth04} to create synthetic light 
curves for the adopted system parameters below. JKTEBOP predicts peak-to-peak 
differences of $>$1\% for inclinations $>$50$\arcdeg$. Given that no 
significant photometric signal is detected above the 0.35\% level, we can safely 
rule out such high inclinations. If we instead use equation 6 from \citet{morris93} 
to predict the amplitude of the reflection effect, we find that the inclination must 
be $i < 58\arcdeg$.  Hence, both light curve modeling and analytic estimates 
require $i \lesssim 50\arcdeg$ otherwise we would see a reflection effect.

\subsection{Joint Analysis}

	We now perform a joint analysis that considers our spectroscopic, 
astrometric, and photometric constraints simultaneously.  Our approach is to 
construct composite binary models by adding two synthetic WD spectra, properly 
weighted by their respective radii using the astrometric parallax constraint.  We 
then simultaneously fit the spectroscopic Balmer line profiles and the broadband 
photometric measurements using the Levenberg-Marquardt algorithm \citep{bedard17, 
kilic20}.  We use de-reddened GALEX FUV and NUV \citep{martin05}, SDSS $ugriz$ 
\citep{abolfathi18}, Pan-STARRS $grizy$ \citep{tonry12}, and UKIDSS $YJ$ photometry 
\citep{lawrence07}. All four atmospheric parameters ($T_{\rm eff,1}$, $\log{g_1}$, 
$T_{\rm eff,2}$, and $\log{g_2}$) are allowed to vary for a given distance.

	Figure \ref{fig:joint} plots the best-fit solution:  a 19,000~K DA WD plus a 
8,000~K DC WD, both with $\log{g}=7$.  We note that a DA+DA solution yields 
comparable parameters, but predicts a strong double-lined H$\alpha$ feature that is 
not observed.  The errors in the joint analysis are dominated by the parallax 
uncertainty.  We estimate 1-$\sigma$ errors of 1000~K in \teff\ and 0.15 dex in 
\logg\ for the DA WD, and 1500~K in \teff\ and 0.25 dex in \logg\ for the DC WD.

	We interpolate the atmospheric parameters through the \citet{althaus13} 
He-core WD tracks to estimate the masses.  Conservatively ignoring the co-variance 
between \teff\ and \logg, we infer $M_{\rm DA, joint} = 0.27^{+0.06}_{-0.02}$~\msun\ 
and $M_{\rm DC, joint} = 0.24^{+0.06}_{-0.04}$~\msun.

	Our mass estimates imply that the cool WD may be less massive than the hot 
WD.  The same result is seen in ZTF J1539+5027 \citep{burdge19}.  This would be 
surprising because one expects the cool WD to have evolved first, and thus be more 
massive than the hot WD.  However the models contain a major uncertainty linked 
to common envelope ejection \citep{li19, li20}.  The thickness of the WD hydrogen 
envelope, assumptions about elemental diffusion and rotational mixing, the presence 
or absence of hydrogen shell flashes, and other issues have significant effects on 
the temperature, radius, and cooling age of a low mass WD \citep{althaus13, 
istrate16}.
	To draw a significant conclusion about the WD masses and their past 
evolution, we need to improve our constraints, i.e.\ with future GW measurements.  
The present constraints are also consistent with J2322+0509 being an equal-mass 
binary.

	Given the mass estimates, we predict that J2322+0509's orbital period is 
shrinking by $\dot{P} = -2.2 \times 10^{-12}$~s~s$^{-1}$ due to GW radiation.  
Measuring $\dot{P}$ is challenging in the absence of a timing signal from eclipses. 
However the phase of the radial velocity orbit also provides a means to measure 
$\dot{P}$.  We simulate radial velocity data to estimate the timing 
constraints.  At the MMT, times are recorded using network time protocol accurate to 
milliseconds, and open-shutter time is accurate to $\sim$0.1~s.  Our current dataset 
has a $\pm$20~s epoch error.  Simulations suggest that 6 contiguous hours on-source 
(4 times our best baseline) with current measurement errors would achieve a 
$\pm10$~s epoch error in a single night.

	We expect J2322+0509's orbital phase to shift by 91~s in 10~yr, and by 364~s 
in 20~yr when LISA is observing.  With $\pm10$~s epoch errors and every-other-year 
observations, we could attain a 5-$\sigma$ measurement of $\dot{P}$ in 14~yr.  The 
two WDs will merge due to GW radiation in $6.7^{+2.8}_{-1.8}$~Myr.

\begin{figure} 
 \includegraphics[width=3.3in, clip, bb=10 215 570 700]{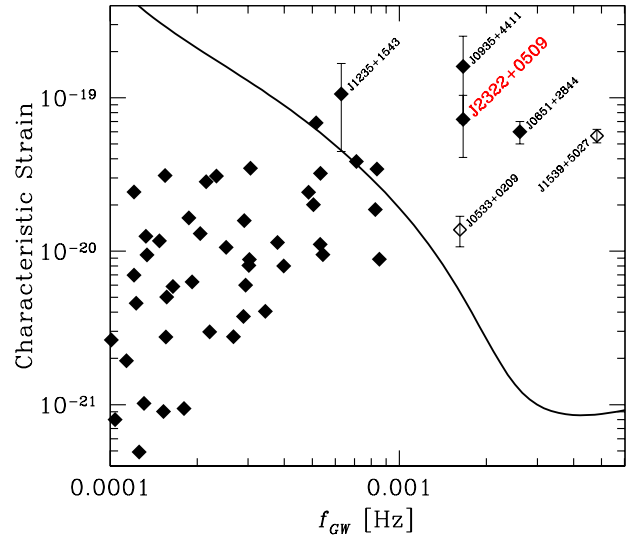}
 \caption{ \label{fig:gw}
	Characteristic strain versus GW frequency for J2322+0509 and other detached 
WD binaries from \citet{brown20} (solid diamonds) and \citet{burdge19, burdge19b} 
(open diamonds).  Solid line is the LISA 4~yr sensitivity curve \citep{robson19}.}
 \end{figure}

\section{Discussion}

	J2322+0509 is a new GW source for LISA to observe.  While the GW community 
may refer to J2322+0509 as a verification binary, optical measurements are critical 
to its scientific understanding.  LISA will continuously monitor the entire sky; 
\citet{shah13} estimate that simply knowing a binary's sky position can improve its 
GW parameter uncertainties by a factor of two.  GW strain depends on binary 
inclination; having an inclination constraint can improve the GW parameter 
uncertainties by a factor of 40 \citep{shah13}.

	Solving the binary mass function for inclination, \begin{equation} \sin{i} = 
k \left( \frac{P}{2\pi G} \right)^{1/3} \frac{(M_1+M_2)^{2/3}}{M_2}, \end{equation}
	our optical measurements of orbital period, velocity semi-amplitude, and 
mass constrain J2322+0509's inclination to $i=27.0 \pm 3.8$ deg.  This 
inclination results in a 2.5 times larger GW strain than if J2322+0509 had 
$i=90\arcdeg$ and were eclipsing.

	Figure \ref{fig:gw} plots the characteristic strain of J2322+0509 relative 
to the 4-year LISA sensitivity curve \citep{robson19}.  Solid diamonds are detached 
WD binaries from the ELM Survey \citep{brown20}, and open diamonds are the two 
detached WD binaries found by \citet{burdge19, burdge19b}.  We calculate strain 
using spectroscopic distance estimates for the sake of consistency, and label only 
those binaries significantly above the 4-year LISA sensitivity curve.

	J2322+0509 has a larger strain than the signature verification binaries 
J0651+2844 and J1539+5027, because it is closer and orientated face-on.  However 
J2322+0509 will be detected at lower SNR because of its lower frequency 
$f_{GW}=2/P=1.66$~mHz.  According to the LISA Detectability Calculator, J2322+0509 
is predicted to have a 4-yr SNR~$\simeq40$ (Q.~S.~Baghi, private communication).  
The same estimator yields SNR=154 for J0651+2844 ($M_1=0.25$ \msun, $M_2=0.5$ 
\msun, $d=1$ kpc, $P=765$ s, $i=86\arcdeg$) as a point of reference.

	Binary population synthesis models predict that most LISA sources will be 
found in the Galactic disk \citep{lamberts19, breivik19}.  J2322+0509 fits that 
picture.  Combining J2322+0509's Gaia parallax and proper motion with our radial 
velocity, corrected for the $6.6\pm0.9$~\kms\ gravitational redshift of the DA 
WD, yields a space motion $(U,V,W) = (-36.1\pm8.0, -20.4\pm4.6, -3.8\pm4.4)$ \kms\ 
with respect to the Sun.  J2322+0509 is located 0.6 kpc below the Galactic plane but 
clearly has the motion of a disk star.

	Interestingly, J2322+0509 is the first He+He WD system among the LISA 
verification binaries.  Population synthesis models predict that 30\% of resolved 
LISA systems will be He+He binaries, 50\% will be He+CO binaries, and 20\% will be 
CO+CO binaries \citep{lamberts19}.  Observationally, all of the detached binaries 
with $f>1$~mHz are He+CO WD binaries, except for J2322+0509.

	Future GW measurements promise to improve our understanding of the WD binary 
systems.  For J1539+5027, \citet{littenberg19} estimate that GW measurements will 
improve its inclination uncertainty by a factor of 5 and its distance uncertainty by 
a factor of 10.  Similar improvements in mass and orbital evolution open the door to 
new tests, for example measuring tidal dissipation in WDs \citep{fuller12, fuller13, 
piro19}.  In order to have the time baseline to measure parameters like $\dot{P}$, 
it is important to continue finding future multi-messenger systems like J2322+0509 
now.

\acknowledgements

	We thank A.\ Milone and Y.\ Beletsky for their assistance with observations 
obtained at the MMT and Magellan telescopes, respectively.  This research makes use 
the SAO/NASA Astrophysics Data System Bibliographic Service.  This work was 
supported in part by the Smithsonian Institution, the NSF under grant AST-1906379,
the NSERC Canada, and by the Fund FRQ-NT (Qu\'ebec).

Based on observations obtained at the Gemini Observatory, which is operated by the 
Association of Universities for Research in Astronomy, Inc., under a cooperative 
agreement with the NSF on behalf of the Gemini partnership: the National Science 
Foundation (United States), National Research Council (Canada), CONICYT (Chile), 
Ministerio de Ciencia, Tecnolog\'{i}a e Innovaci\'{o}n Productiva (Argentina), 
Minist\'{e}rio da Ci\^{e}ncia, Tecnologia e Inova\c{c}\~{a}o (Brazil), and Korea 
Astronomy and Space Science Institute (Republic of Korea).

\facilities{MMT (Blue Channel spectrograph), Magellan:Clay (MagE spectrograph), 
	Gemini:North (GMOS spectrograph)}

\software{IRAF \citep{tody86, tody93}, RVSAO \citep{kurtz98}, 
	JKTEBOP \citep{southworth04}}

\begin{deluxetable}{lc}
	\tablecolumns{2}
	\tablewidth{0pt}
	\tablecaption{J2322+0509 System Parameters\label{tab:param}}
	\tablehead{ \colhead{Parameter} & \colhead{Value} }
	\startdata
\sidehead{Astrometric (Gaia DR2)}
RA (J2000)		& 23:22:30.203 \\
Dec (J2000)		& +5:09:42.061 \\
Plx (mas)		&  1.287  $\pm$ 0.283 \\
$\mu_{\rm RA}$ (\mas)	& 11.094  $\pm$ 0.428 \\
$\mu_{\rm Dec}$ (\mas)	& -7.119  $\pm$ 0.336 \\
\sidehead{Photometric (SDSS DR14)}
$u$ (mag)		& 18.853  $\pm$ 0.028 \\
$g$ (mag)		& 18.589  $\pm$ 0.018 \\
$r$ (mag)		& 18.906  $\pm$ 0.015 \\
$E(B-V)$ (mag)		& 0.062 \\
\sidehead{Spectroscopic (MMT+Magellan)}
$P$ (s)			& 1201.4  $\pm$  5.9 \\
$k$ (\kms)		&  148.6  $\pm$  6.3 \\
$\gamma$ (\kms)		&   -4.8  $\pm$  4.2 \\
$T_0$ BJD$_{\rm TDB}$ (d)	& 2458764.685470  $\pm$ 0.000228 \\
$T_{\rm eff,DA,spec}$ (K)	&  19160  $\pm$ 270 \\
$\log{g}_{\rm \,DA,spec}$ (\cms)&   7.17  $\pm$ 0.04 \\
\sidehead{Joint Analysis (Adopted)}
$d_{\rm plx}$ (kpc)		&   0.76  $\pm$ 0.17 \\
$T_{\rm eff,DA}$ (K)		&  19000  $\pm$ 1000 \\
$\log{g}_{\rm \, DA}$ (\cms)	&    7.0  $\pm$ 0.15 \\
$T_{\rm eff,DC}$ (K)		&   8000  $\pm$ 1500 \\
$\log{g}_{\rm \, DC}$ (\cms)	&    7.0  $\pm$ 0.25 \\
\sidehead{Derived}
$M_{\rm DA}$ (\msun)	& $0.27^{+0.06}_{-0.02}$ \\
$M_{\rm DC}$ (\msun)	& $0.24^{+0.06}_{-0.04}$ \\
$a$ (\rsun)		& $0.194 \pm 0.007$ \\
$i$ (deg)		& $ 27.0 \pm 3.8$ \\
	\enddata
\end{deluxetable}

\clearpage

\end{document}